\documentstyle[aasms4]{article}

\def\be{\begin{equation}}
\def\en{\end{equation}}

\begin{document}
\begin{titlepage}
\baselineskip = 25pt
\begin{center}
{\Large\bf LEARNING FROM OBSERVATIONS OF THE
MICROWAVE BACKGROUND AT SMALL ANGULAR SCALES}

\vspace{.5 cm}
{\bf $^{1}$D. S\'aez 
and $^{2}$J.V. Arnau}\\
\small
$^{1}$Departamento de Astronom\'{\i}a y 
Astrof\'{\i}sica. Universidad de Valencia.\\
46100 Burjassot (Valencia), Spain.\\
$^{2}$Departamento de Matem\'atica Aplicada. Universidad de Valencia.\\
46100 Burjassot (Valencia), Spain.\\
\footnotesize
e-mail: diego.saez@uv.es, jose.arnau@uv.es\\
\end{center}

\vspace {2. cm}
\normalsize
\begin{abstract}

In this paper, we focus our attention on  
the following question: How well can we
recover the power spectrum of the cosmic microwave background	
from the maps of a given experiment?. Each experiment is described by a
a pixelization scale, a beam size, a noise level and a sky coverage.
We use accurate numerical simulations of the microwave sky
and a cold dark matter model for structure formation 
in the universe. 
Angular scales smaller than those of previous simulations are 
included.
The spectrum obtained from the simulated maps
is appropriately compared with the theoretical one.
Relative deviations between these spectra are estimated.
Various contributions to these deviations are analyzed.
The method used for spectra comparisons is discussed.

\end{abstract}

{\em Subject headings:} cosmic microwave background---cosmology:
theory---large-scale structure of the universe

\end{titlepage}

\section{INTRODUCTION}

This paper is devoted to the estimation of the 
cosmic microwave background (CMB) anisotropies
from the maps of a given experiment. The goodness of a certain experiment
can be tested by using numerically simulated maps 
similar to those of the chosen experiment.
Simulations require a theoretical 
model for structure formation in the universe. Since we are
particularly interested in the analysis of some experiment
features,
only simulations based on the theoretical model 
of Sec. 2 are considered. Other models will be studied elsewhere.

From the spectrum of the CMB anisotropies 
corresponding to a theoretical
model 
--defined by the quantities 
$C_{\ell} \equiv \sum_{m = -\ell}^{m = \ell} |a_{\ell m}|^2 
/(2\ell + 1) $--
and the features of a certain experiment, 
we can build up simulated maps.  From these maps, the initial spectrum	
can be partially recovered. The final spectrum deviates from the
initial one. The main goal of this paper is the study of various 
important 
contributions to these deviations. The following facts are taken
into account:
(1) The existence of an angle, $\theta_{min}$, 
defining a regular network on the sky (pixelization),  
(2) the use of a Gaussian antenna with a  
full-width at half-maximum angle $\theta_{_{FWHM}}$. 
This angle does not fix the pixelization scale. 
Angles $\theta_{min}$ and $\theta_{_{FWHM}}$
are independent in spite of the fact that they usually take on
similar values; for example, in the case of an ideal 
detector with $\theta_{_{FWHM}}=0$, 
a nonvanishing $\theta_{min}$ value 
could be required by the observational strategy,
(3) the fact that only one realization of the
CMB sky is available from our position in the universe. 
This is the cosmic deviation (usually described by the so-called
cosmic variance),
(4) the partial coverage 
of the unique available CMB realization. 
The size --area-- and the shape --including possible holes-- of the 
observed regions are important
(Scott, Srednicki \& White 1994) and,  
(5) the simultaneous existence of white noise and a partial sky coverage.

The method used in order to recover the angular power spectrum from 
the simulated maps is described and analyzed along the paper.
This method appears to be an interesting alternative to 
usual methods based on Fourier analysis, which require the use
of small maps. See S\'aez, Holtmann \&
Smoot (1996) and Hobson \& magueijo (1996) for interesting applications 
of Fourier tecniques to the generation and analysis of appropriate
patches of the CMB sky. 
The proposed method 
is particularly appropriate in the case of maps covering a 
great region of the sky. These maps can be studied as 
a whole. They can include holes.
Other interesting features of this method are
pointed out in Secs. 4, 5 and 6.

The contamination produced by the Milky Way
and other galaxies is not studied in this paper. This is not a 
dramatic restriction for experiments with appropriate  
frequencies greater than 
$\sim 80 \ GHz$ and smaller than  $\sim 120 \ GHz$.
For these frequencies, 
the emissivities from the Milky Way
and the extragalactic foregrounds are minimum. At 
small-intermediate angular scales,  
temperature fluctuations produced by these emissions are expected to be
near $10^{-6}$. Observations in the above frequencies are only 
feasible in the case of satellite experiments.
For ground-based and balloon-borne
experiments, the atmosphere prevents observations at frequencies
lying between $80 \ GHz$ and $120 \ GHz$. 
Observations must be carried out at
frequencies much smaller than $80 \ GHz$ and, then, the
contributions of the galaxy and the extragalactic sources must be 
carefully subtracted in a model dependent way
(Tegmark \& Efstathiou 1995, Dodelson 1995). 

This paper gives a partial answer to the following
question: What can we learn from satellite observations
($80 \ GHz < \nu < 120 \ GHz$) 
with high sensitivities and coverages plus accurate
simulations of the microwave sky?. Discussing  
this question, we are contributing to 
justify the required observations and simulations, which
have a high cost in all the senses.

\section{THE MODEL}

All the simulations presented in this paper 
correspond to an unique cold dark matter
model for large scale structure formation.
This model is defined by the following assumptions: 
(i) after standard recombination
and decoupling, no reionizations modified the anisotropies of the
CMB, (ii) the background is
flat ($\Omega_{0}=1$), (iii) the cosmological constant vanishes, 
(iv) scalar fluctuations are Gaussian and their spectrum
is scale-invariant, and (v) tensor fluctuations are absent.
In this model,
the resulting anisotropy depends on the
evolved spectrum of the scalar modes, which depends on the
primordial spectrum and the quantities involved in the
transfer function.
This function involves the
density parameter of the baryonic matter $\Omega_{_{B}}$
and the reduced Hubble constant $h$. 
The values of the parameters $h$ and $\Omega_{_{B}}$ are assumed to be
$1/2$ and $0.03$, respectively. 

The angular power spectrum of the CMB anisotropy is usually
normalized by the rms quadrupole 
produced by scalar modes. In this paper, normalization is based on  
the estimator $Q_{rms-PS}$, which is obtained by fitting the observed 
temperature fluctuations in the case of a scale-invariant
primordial density power spectrum.
In the absence of tensor modes,
experiments measuring at large angular scales --as COBE (Smoot
et al. 1992, Bennet et al. 1992, Wright et al. 1992) 
and TENERIFE (Hancock et al. 1994) 
lead to estimations of $Q_{rms-PS}$. In this paper, the
$C_{\ell}$ coefficients
have been taken from Sugiyama (1995) and
renormalized according to the four year  
COBE data ($Q_{rms-PS} \simeq 18 \ \mu K$, G\'orski et al. 1996).

Let us give some basic definitions and comments, which are
used below:
The autocorrelation function can be defined as 
$C(\theta)=C_{\sigma=0}(\theta)$, where
\be 
C_{\sigma}(\theta)=\left \langle 
\left( \frac {\delta T}{T} \right)_{\sigma} ({\vec {n}}_{1})
\left( \frac {\delta T}{T} \right)_{\sigma} ({\vec {n}}_{2})
\right \rangle \ .  
\en
The angle between the unit vectors $\vec {n}_{1}$ and $\vec {n}_{2}$
is $\theta$. 
The quantity $(\frac {\delta T}{T})_{\sigma} ( \vec {n})$
is the temperature contrast in the direction $ \vec {n} $ after
smoothing with a Gaussian beam described by
$\sigma=0.425 \theta_{_{FWHM}}$. The
angular brackets stand for a mean on many CMB realizations.
Function $C_{\sigma}(\theta)$ can be expanded in the following
form:
\be
C_{\sigma}(\theta)= \frac {1}{4 \pi} \sum_{\ell =2}^{\infty}
(2\ell +1)C_{\ell} P_{\ell}(\cos {\theta}) e^{- (\ell + 0.5)^{2}
\sigma^{2}}.
\en
From Eq. (2) one easily obtains the relation: 
\be
C_{\ell}(\sigma ) = e^{- (\ell + 0.5)^{2}
\sigma^{2}} C_{\ell}
=\frac {32 \pi^{3}} {(2 \ell +1)^{2}}
\int_{0}^{\pi} C_{\sigma}
(\theta) P_{\ell} (\cos {\theta}) sin \theta d \theta \ .
\en

\section{EXPERIMENTS AND SPECTRA}

In order to present a systematic study of the   
uncertainties in the measurement
of the CMB spectrum, three
auxiliary experiments are considered. 
The main features of these experiments
are now listed:
  
Experiment (A):  $\theta_{min} \neq 0$, 
$\theta_{_{FWHM}}=0$, no noise, and partial coverage.

Experiment (B): $\theta_{min} \neq 0$, 
$\theta_{_{FWHM}} \neq 0$, no noise, and partial coverage.

Experiment (C):  $\theta_{min} \neq 0$, 
$\theta_{_{FWHM}} \neq 0$, uncorrelated noise, and partial coverage.

Experiment (A) corresponds to a perfect detector measuring 
temperatures at the nodes
of a discrete regular grid covering a part of the sky.
Some aspects of this experiment
can be studied without simulations. The method used in order
to extract the spectrum from
given maps of the CMB anisotropy is as follows: 
the autocorrelation function $C( \theta)$ is estimated from 
Eq. (1) and, then,
Eq. (3) is used in order to get $C_{\ell}$ quantities.
Many pairs of directions 
($\vec {n}_{1}$, $\vec {n}_{2}$) forming a given angle $\theta$
are randomly placed on the available maps in order to 
perform the average involved in Eq. (1). 
The number of pairs is experimentally fixed (It is verified that
a number of independent pairs greater than the chosen one does not
lead to a better estimate of the average).
An accurate determination
of $C( \theta)$ requires various full realizations of the CMB
sky. If these realizations are not available, there are errors
in the resulting autocorrelation function and, 
consequently, there are errors
in the $C_{\ell}$ coefficients given by Eq. (3).
Since only a realization of the CMB sky is available, there is an unavoidable 
indetermination in the CMB spectrum 
(cosmic uncertainty). Unfortunately, a
full coverage of our CMB sky is not available. Up to date,	
only small regions of the sky have been observed (except
in the case of large angular scales). Future 
satellite experiments could give a more complete coverage, but
contaminations due to the Milky way and other galaxies could either
require or suggest the rejection of large regions in some maps.

Let us now consider some uncertainties in $C( \theta)$ appearing
as a result of the existence of both an angle $\theta_{min}$
separating neighboring nodes of the grid and  
an angle $\theta_{max}$ associated to a partial coverage of the sky.
In other words, 
from Eq. (1), the function $C( \theta)$ can be only obtained in a certain
interval ($\theta_{min}$,$\theta_{max}$). For angles smaller than 
$\theta_{min}$, the map have not any information. As a result of
partial coverage, the great number of 
($\vec {n}_{1}$, $\vec {n}_{2}$) pairs required by Eq. (1)
is only feasible for angles smaller than a 
certain $\theta_{max}$.
This means 
that the maps have not information for too small ($\theta < \theta_{min}$)
and too large ($\theta > \theta_{max}$) angular
scales. 
This discussion holds for both observation maps and simulated ones. 
Even if the temperatures have been accurately measured
in the grid nodes, the integration involved in Eq. (3)
can only be extended to the interval ($\theta_{min}$,$\theta_{max}$)
--not to the interval ($0$,$\pi$)-- and, 
consequently, this integration leads to 
$C_{\ell}$ values different from the theoretical ones.
These values define a certain spectrum which is 
is hereafter called the {\em modified 
spectrum} to be distinguished from the {\em true spectrum} 
corresponding to the theoretical model under consideration.
The true CMB spectrum cannot be directly obtained from Eq.\ (1) 
--namely, from the definition of the autocorrelation function-- 
as a result of intrinsic limitations in the
maps and, consequently, we must be cautious with any 
indirect mathematical method
creating (modifying) information outside (inside) the interval  
($\theta_{min}$,$\theta_{max}$) to recover the
true $C_{\ell}$ quantities. Further discussion about this point
is given in Sec. 6. 

The modified spectrum obtained from the maps is not to be compared
with the true spectrum but with the {\em theoretical modified 
spectrum} obtained as follows: first,  Eq. (2)  and 
the $C_{\ell}$ coefficients of the assumed 
model --for $2 \leq \ell \leq 1100$-- are used 
in order to get the function $C(\theta)$
in the interval $(0, \pi )$; afterwards,
the values of $C(\theta)$ in the interval 
($\theta_{min}$,$\theta_{max}$) and Eq.\ (3) are used to get   
the theoretical modified 
$C_{\ell}$ quantities for $40 \leq \ell \leq 1000$.

The theoretical modified spectra
corresponding to several values of $\theta_{min}$ and $\theta_{max}$
are displayed in Fig.\ 1, where
the continuous line corresponds to the true $C_{\ell}$ coefficients
of the chosen model (see Sec. 2).
In the top panel, the value $\theta_{max}=9^{\circ}$ is
fixed, while the angle $\theta_{min}$ takes on the values
$2.5^{\prime}$ (pointed line), $5^{\prime}$
(dashed line), and $10^{\prime}$ (pointed-dashed line). 
From $\ell=40$ to $\ell=1000$, 
the true $C_{\ell}$ quantities can be only recovered with 
high accuracy for
very small values of $\theta_{min}$. 
In the bottom panel, the value $\theta_{min}=5^{\prime}$ is
fixed, while the angle $\theta_{max}$ takes on the values
$4.5^{\circ}$ (pointed line), $9^{\circ}$
(dashed line), and $18^{\circ}$ (pointed-dashed line). 
From $\ell=40$ to $\ell \sim 280$, the effect of a varying $\theta_{max}$
is significant; however, for $\ell>280$, this effect
becomes very small. 

Fig.\ 1 shows that, for large angles, the theoretical modified
spectrum is affected by pixelization
($\theta_{\min}$ value). This fact does not make
difficult the evaluation of theories because the 
theoretical modified spectrum is to be compared with 
the modified spectrum obtained from the maps, which is affected 
by pixelization in the same way.   

Experiment (B) admits the same discussion as the previous one. Formulae
are the same but the beam size does not vanish.  
Two beams have been considered in this paper:  $\sigma=2.125^{\prime}$ 
($\theta_{_{FWHM}}=5^{\prime}$) and 
$\sigma=4.25^{\prime}$
($\theta_{_{FWHM}}=10^{\prime}$). 

The  
$C_{\ell}(\sigma)$ quantities obtained from simulated maps
--corresponding to given values of $\theta_{min}$ and $\sigma$-- 
must be compared with the theoretical modified quantities
corresponding to the same $\theta_{min}$ and $\sigma$ values. 
The angle $\theta_{max}$ must be 
compatible with the simulation coverage.
If the modified spectra are used for comparisons, 
the uncertainties due to $\theta_{min}$ and $\sigma$
become separated from other uncertainties due to partial coverage,
noise and foregrounds. Comparisons with the true $C_{\ell}(\sigma)$
do not lead to this separation. 
Other uncertainties in the determination of the CMB 
spectrum can be analyzed by using simulations (see Sec. 5).

Our estimation of the $C_{\ell}(\sigma)$ 
quantities is directly based on the definition 
of the autocorrelation function (namely,
on Eq. (1)). As stated before, pairs    
($\vec {n}_{1}$, $\vec {n}_{2}$) are appropriately located on the maps.
This method is so simple and direct that: 
(1) It applies to the case of any
extented map including holes in a natural way and, 
(2) the analysis of errors in spectra estimates is	
very simple. Errors seem to be associated to intrinsic
limitations of the maps (pixelization, partial coverage et cetera).
These limitations lead to problems with the 
location of pairs  ($\vec {n}_{1}$, $\vec {n}_{2}$)
(see Sec. 6).

\section{SIMULATIONS}

Our numerical simulations are extended to 
$40^{\circ} \times 360^{\circ}$
regions of the sky. These regions are assumed to be uniformly
covered and, consequently, the angle 
$\theta_{min}$ --giving the separation between neighboring
points- defines the grid of the simulated maps. 

Simulations are based on the expansion:
\be
\frac {\delta T}{T}(\theta,\phi)=\sum_{\ell =1}^{\ell_{max}}
\sum_{m=-\ell}^{m=+\ell}
a_{\ell m}Y_{\ell m}(\theta,\phi),
\en
with $\ell_{max} = 1100$. 
The  $a_{\ell m}$ coefficients have been generated as statistically
independent random numbers with
variance $\langle  \mid a_{\ell m} \mid ^{2} \rangle = C_{\ell}
e^{(- \ell + 0.5)^{2} \sigma^{2}}$
and zero mean.
The spherical harmonics have been carefully calculated.  
These simulations include scales smaller than those 
considered in previous ones
(see Hinshaw, Bennett \& Kogut 1995,
Kogut, Hinshaw \& Bennett 1995, Jungman et al. 1995). 
The small $\theta_{min}$
and $\sigma$ values considered in our simulations (see below) 
require the use of large $\ell$ values giving information about 
small angular scales. 

In a IBM 30-9021 VF, the CPU cost is $\sim 11$ hours per simulation.
The CPU cost for simulations of the full sky has been estimated to 
be  
$\sim 50$ hours.

\section{RESULTS FROM SIMULATIONS}

We begin with the experiment (B) for several coverages.

Three $40^{\circ} \times 360^{\circ}$
simulations (C1 coverage) cover a surface of 43200 square degrees.
This is an area slightly greater than that of the full sky
(41253 square degrees). The Milky Way mainly contaminates a band
of $40^{\circ} \times 360^{\circ}$
and, consequently, taking a conservative point of view, we
could try to extract the spectrum of the CMB anisotropy by
using two simulations (coverage C2), 
namely, a total area of 28800 square degrees.

In the absence of noise, holes and other errors in 
the estimation of the spectrum, coverage C1 
should lead to
a realization of the cosmic uncertainty; in other words, three bands should
lead to results comparable to those of a full realization of
the CMB sky, at least, for large $\ell$ values. 

The following coverages are also considered:  a full band (C3), 
a band with four $40^{\circ} \times 40^{\circ}$ separated
squared holes (C4), and an unique $40^{\circ} \times 40^{\circ}$
squared region extracted from a band (C5). Results from 
cases C1 -- C5 give interesting 
information about the uncertainties in the resulting
spectrum appearing as a result of the coverage features.

The left and right top panels of 
Fig.\ (2) shows the power spectrum obtained from two
different C1 realizations.
The beam size is
$\sigma = 2.125^{\prime}$ and
$\sigma =4.25^{\prime}$ in the right and left panels,
respectively. In both cases, 
the grid is defined by the angle $\theta_{min}=5^{\prime}$. 
Solid lines are
the theoretical modified spectra corresponding to the chosen values of 
$\sigma$ and $\theta_{min}$. 

In order to measure the deviations between the theoretical
and simulated spectra, the following quantities are calculated
and presented in Table 1 : The mean, $M1$, of the quantities 
$0.69 \ell (\ell +1)C_{\ell}(\sigma) 
\times 10^{10}$ (column 3), the mean, $M2$, of the 
differences
between theoretical and simulated values of these quantities
(column 4), the mean, $MA$, of the absolute value of these differences 
(column 5),
and the typical deviation, $\Sigma$, of the differences 
of column 4 (column 6). 
These
quantities are estimated in appropriate $\ell$ intervals  
(column 7). 

The intermediate and bottom panels of Fig.\ 2 have the same structure as 
the corresponding top panels, but in the 
intermediate (bottom) panel, the dashed line corresponds
to the C3 (C5) coverage. 
From these panels and Table 1, 
it follows that, as expected, the existence of holes in 
$40^{\circ} \times 360^{\circ}$
bands and, in general, the incompleteness of the
sky coverage lead to significant uncertainties in the 
spectrum. 

Simulations have showed that 
the simulated spectra essentially oscillate around the
modified theoretical one. This is a very good news in order
to stablish comparisons with theoretical models
(see Sec. 6).
The presence of oscillations is pointed out by the relation
$\mid M1 \mid < M2$, which is satisfied for every coverage
(see Table 1). The fact that $M1$ is always negative --except
in the case of the last entry of Table 1-- indicates
the existence of a systematic error for large $\ell$ values.
The same indication is obtained from the top panels of Fig.\ 2,
where it can be seen that, for large $\ell$ values, the curves
corresponding to simulations lie slightly below the theoretical ones
(see Sec. 6 for an interpretation of this fact).

Finally, the experiment (C) has been considered in order to take
into account the possible combined effect of uncorrelated noise and
partial coverages (for $\ell > 40$). 
It has been verified that this combined effect is 
negligible for the coverages C1 - C5 and a noise level of 
$27.3 \ \mu K$. Maps 
involving pure white noise lead to no correlations ($C(\theta)=0$)
for the ensemble, but pure white noise
can give nonvanishing $C_{\ell}$ coefficients in the case of
an unique sky realization (cosmic variance) or in the case
of partial coverage.
In order to test the importance of the combined effect
of white noise and partial coverage, a
$40^{\circ} \times 40^{\circ}$ map has been built up. This map only
involves
pure uncorrelated noise at a level of $27.3 \ \mu K$. The resulting
$C_{\ell}$ 
quantities have been extracted as in any other case. The
values of $0.69 \ell (\ell+1) C_{\ell} \times 10^{10}$ 
have appeared to be smaller than $10^{-2}$ for any scale. These
values are much smaller than those of Figs.\ 1 and 2 
(order unity),
which correspond to cosmological signals. In conclusion, 
the presence of white noise at a level of $27.3 \ \mu K$
can only be important either in the
case of coverages much smaller than C5 or in the case
$\ell < 40$. According to our
expectations,
it has been verified that the smaller the coverage, the greater the 
relevance of uncorrelated noise.

\section{CONCLUSIONS AND DISCUSSION}

The use of modified spectra (see Sec. 3)
allows us to separate the effects of 
smoothing and pixelization from other effects.
The form of the modified spectra depends on 
$\theta_{min}$, $\theta_{max}$ and $\sigma$. 
These spectra are to be compared with
those extracted from observations or simulations. 
The main problems with the estimate of the 
modified spectra are now discussed.

In order to obtain $C_{\sigma}(\theta)$
from a given map, many pairs ($\vec {n}_{1} , \vec {n}_{2}$) are
randomly located on the map. Direction $\vec {n}_{1}$ can be randomly placed
on a node of the grid, but then, for a given $\theta$, direction
$\vec {n}_{2}$ does not point towards another node. This mean that
the temperature in the direction $\vec {n}_{2}$
is not known and, consequently, it must be estimated by using
interpolations in the grid. This is a
mathematical method introducing wrong information. 
Let us discuss this point in more detail. 
In order to get true temperatures outside the nodes,
we would need a greater resolution in the experiment (or simulation)
and greater $\ell$ values (namely, physical 
improvements). The fictitious values generated
by interpolation can produce an error whose form is
not known from theory.

It has been 
verified that our estimation of the modified spectrum
is good for angles lying in the
interval ($\theta_{min}$, $\theta_{max}$), but 
a small systematic
error (see also Sec. 5) seems to appear as a result 
of the mentioned interpolation.  It
decreases as 
$\theta$ increases; hence, it is more important
for large $\ell$ values. This error remains small up to $\ell=1000$
(see Fig.\ 2). In the case of large $\theta$ values, 
great coverages avoid problems with the location of pairs
($\vec {n}_{1} , \vec {n}_{2}$). According to Sec. 3,
full sky coverages could be particularly important
in the case $\ell < 280$.

Sky coverage is important. Figure 2 and Table 1 show the deviations between
theoretical and simulated spectra for several coverages. As
expected, the 
greater the coverage, the smaller the deviations
(Scott, Srednicki \& White 1994).  Simulations have showed some 
relevant
features of these deviations. It is noticeable that 
an important part of the deviations shows an oscillatory character
around a curve very close to the theoretical one. It can be seen that,
even for the $C5$ coverage, the best fitting to the 
oscillating values is a curve very close to the theoretical one.
This fact enhances the interest of moderated coverages 
as C5 (bottom panels of Fig.\ 2), which lead to
a spectrum very similar to that of C1 coverages (top panels
of Fig.\ 2), at least, for $\ell > 200$ and 
after removing oscillatory deviations.
This is true even in the case of the bottom right panel of Fig.\ 2,
where  one of the most oscillating C5 realizations
has been showed (realizations of this type
are not abundant). 
More abundant realizations oscillate 
as in the left bottom panel. These facts --pointed out by our accurate
simulations-- enhance the interest of 
C5 and similar coverages; in particular, if it is taking into account that
maps with these features 
can be obtained by measuring in selected regions with small contamination. 

For $\ell > 40$,
uncorrelated noise does not appear to be relevant for the
coverages considered in this paper; however, this noise could
be important in the case of smaller coverages.

If the normalization of the $C_{\ell}$ coefficients is performed
according to other estimations of $Q_{rms-PS}$ 
(Smoot et al (1992), Bennet et al (1994),
G\'orski et al (1994), or future estimates),
the above $C_{\ell}$ quantities 
would appear either reduced or magnified
by the factor $\sim (Q_{rms-PS}/18)^{2}$; nevertheless,
the main conclusions of this paper would remain unaltered
because they do not depend on normalization.

\vspace{1 cm}

\noindent
{\it\bf Acknowledgments}. This work has been 
supported by the  Generalitat Valenciana (project GV-2207/94).
Calculations were carried out in a IBM 30-9021 VF at the Centre de
Inform\'atica de la Universidad de Valencia. 

\newpage

\begin{table}
\begin{center}
TABLE 1\\
COMPARING THEORETICAL AND SIMULATED SPECTRA\\
 \begin{tabular}{ccccccc}\\
\hline
\hline
$\theta_{_{FWHM}}$ & coverage  & M1  & M2 & MA & $\Sigma$ & $\ell$-interval\\
\hline
$10^{\prime}$ & C1 & $1.07$ &  $-2.77 \times 10^{-2}$ &
      $4.66 \times 10^{-2}$ & $1.97 \times 10^{-3}$ & $40 - 1000$\\
$10^{\prime}$ & C2 & $1.07$ &  $-2.72 \times 10^{-2}$ &
      $5.36 \times 10^{-2}$ & $2.21 \times 10^{-3}$ & $40 - 1000$\\
$10^{\prime}$ & C3 & $1.07$ &  $-3.04 \times 10^{-2}$ &
      $7.85 \times 10^{-2}$ & $3.24 \times 10^{-3}$ & $40 - 1000$\\
$10^{\prime}$ & C4 & $1.07$ &  $-3.69 \times 10^{-2}$ &
      $1.00 \times 10^{-1}$ & $4.36 \times 10^{-3}$ & $40 - 1000$\\
$10^{\prime}$ & C5 & $1.07$ &  $-3.73 \times 10^{-2}$ &
      $1.85 \times 10^{-1}$ & $7.2 \times 10^{-3}$ & $40 - 1000$\\
$5^{\prime}$ & C1 & $1.23$ &  $-3.80 \times 10^{-2}$ &
      $6.91 \times 10^{-2}$ & $2.74 \times 10^{-3}$ & $40 - 1000$\\
$5^{\prime}$ & C2 & $1.23$ &  $-4.06 \times 10^{-2}$ &
      $7.86 \times 10^{-2}$ & $3.15 \times 10^{-3}$ & $40 - 1000$\\
$5^{\prime}$ & C3 & $1.23$ &  $-4.21 \times 10^{-2}$ &
      $1.01 \times 10^{-1}$ & $3.98 \times 10^{-3}$ & $40 - 1000$\\
$5^{\prime}$ & C4 & $1.23$ &  $-4.30 \times 10^{-2}$ &
      $1.25 \times 10^{-1}$ & $5.04 \times 10^{-3}$ & $40 - 1000$\\
$5^{\prime}$ & C5 & $1.23$ &  $-5.44 \times 10^{-2}$ &
      $2.70 \times 10^{-1}$ & $1.10 \times 10^{-2}$ & $40 - 1000$\\
$5^{\prime}$ & C1 & $2.20$ &  $1.73 \times 10^{-2}$ &
      $5.49 \times 10^{-2}$ & $3.80 \times 10^{-3}$ & $40 - 315$\\
\hline
\multicolumn{7}{c}{}\\
\end{tabular}
\end{center}
\end{table}

\newpage

{\Large\bf References}
\\
Bennett et al. 1992, ApJ, L7\\
Bennett, C.L. et al. 1994, ApJ, 436, 423\\
Dodelson, S., 1995, submitted to ApJ, astro-ph/9512021\\
G\'orski, K.M., et al., 1994, ApJ, 430, L89\\
G\'orski, K.M., et al., 1996, Astro-ph/9601063, submitted to
ApJ. Letters\\
Hancock, S., et al. 1994, Nature, 367, 333\\
Hinshaw, G., Bennett, C.L. \& Kogut, A., 1995, ApJ, 441, L1\\
Hobson \& magueijo, 1996, Astro-ph/9603064\\
Jungman, G., Kamionkowski, M., Kosowsky, A. \& Spergel, D.N.,
1995, Astro-ph/9512139\\
Kogut, A., Hinshaw, G. \& Bennett, C.L., 1995, ApJ, 441, L5\\
S\'aez, D., Holtmann, E. \& Smoot, G. F., 1996,
astro-ph/9606164, accepted in ApJ.\\
Scott, D., Srednicki, M. \& White, M., 1994, ApJ, 421, L5\\
Smoot et al. 1992, ApJ, 396, L1\\
Sugiyama, N., 1995, ApJ Supplement, 100, 281\\
Tegmark, M. \& Efstathiou, G., 1995, submitted to MNRAS,
astro-ph/9507009\\
Wright, E.L. et al. 1992, ApJ , 396, L13\\

\newpage

\begin{center}
{\bf Figure Captions}
\end{center}

\noindent
{\bf Fig.\ 1.} Each panel shows the quantity
$0.69 \ell (\ell + 1) C_{\ell}(\sigma) \times 10^{10}$ as a function of 
log($\ell$) in various cases. In both panels, the solid
line corresponds to $\sigma = 0, \theta_{min}=0$ and, 
$\theta_{max}= \pi$ (true 
$C_{\ell}$ quantities). In
the top panel:  $\theta_{max}=9^{\circ}$, 
$\sigma=0$, and curves are labelled by $\theta_{min}$ in minutes.
In the bottom panel: $\theta_{min}=5^{\prime}$, 
$\sigma=0$, and curves are characterized by  $\theta_{max}$ in degrees.
\vskip 0.5cm
\noindent

\noindent
{\bf Fig.\ 2.} Same as Fig.\ 1 in other cases. 
Left (right) panels 
correspond to $\theta_{_{FWHM}}=
10^{\prime}$ ($\theta_{_{FWHM}}=5^{\prime}$). In all these panels,
$\theta_{min}=5^{\prime}$ and $\theta_{max}=9^{\circ}$. 
The continuous line shows the
modified theoretical spectrum and the dashed line exhibits
that extracted from simulations. Top, intermediate, and 
bottom panels correspond to the C1, C3, and C5 coverages
defined in the text, respectively.
\vskip 0.5cm
\noindent

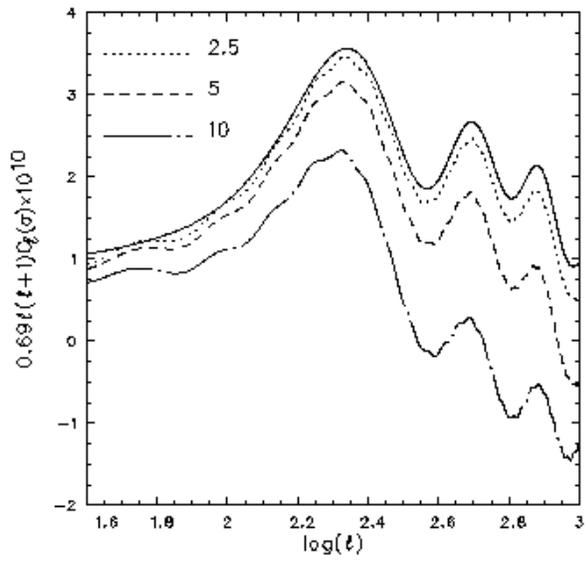
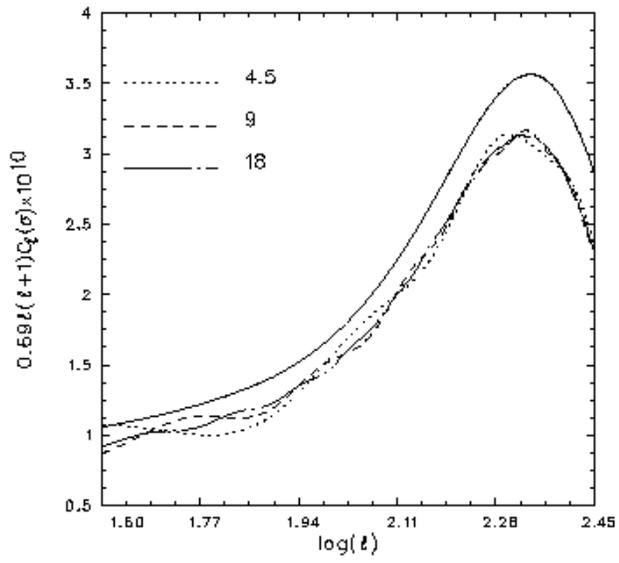
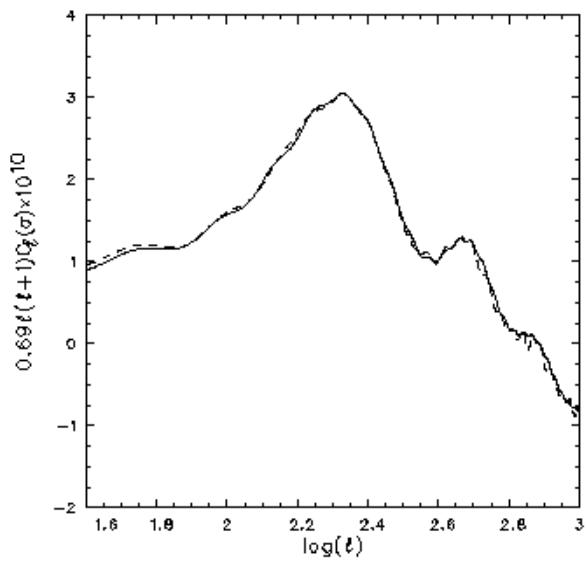
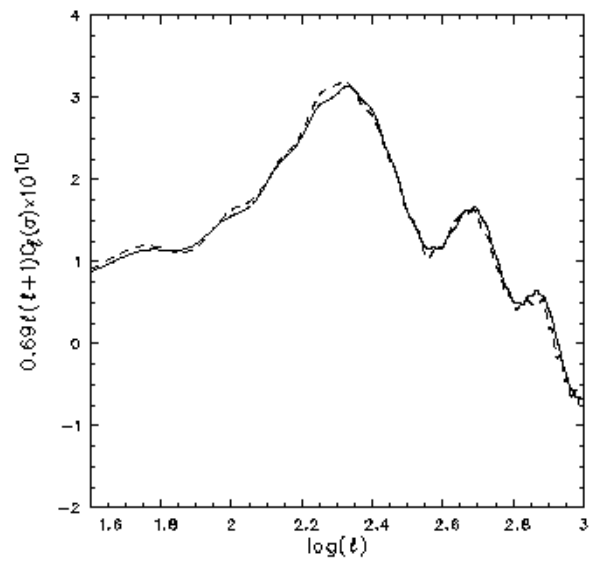
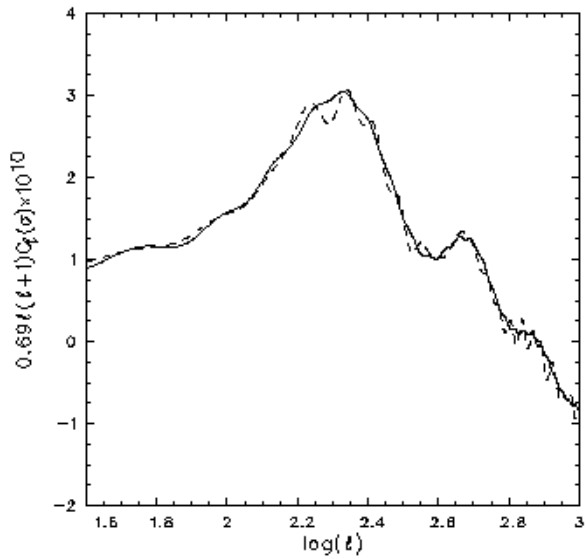
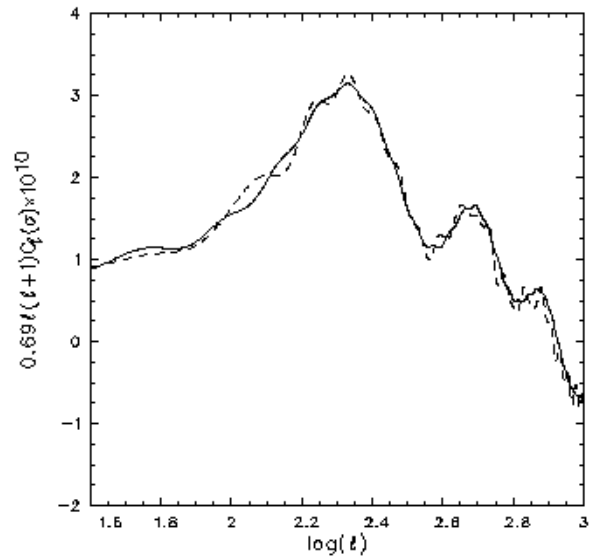
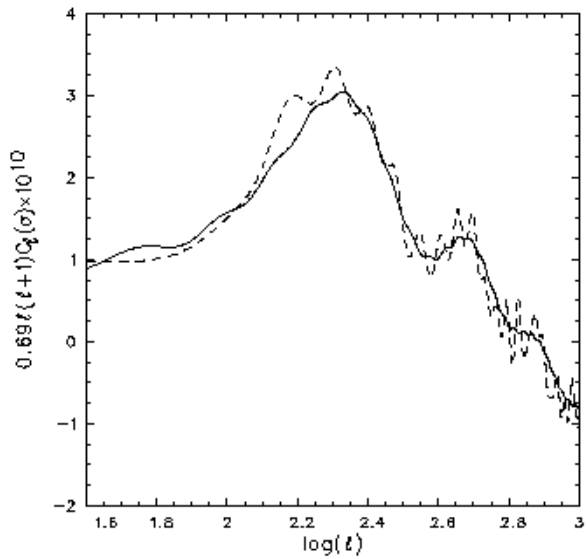
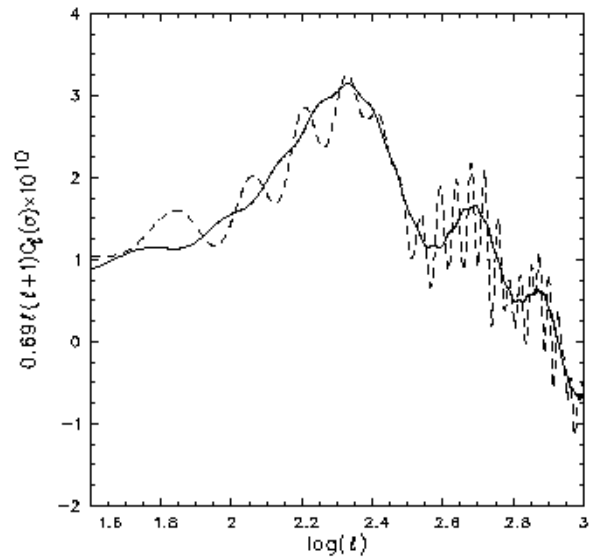
\end{document}